\documentclass{aa}
\usepackage{epsfig}
\usepackage{natbib}
\usepackage[english]{babel}
\usepackage[utf8]{inputenc}
\usepackage[labelfont=up]{subcaption}
\usepackage[x11names, table]{xcolor}
\usepackage{float}
\usepackage{mathtools}
\usepackage[top=1.in, bottom=1.in, left=0.5in, right=0.5in]{geometry}
\usepackage{graphicx}
\usepackage{lipsum}
\usepackage{bm}
\usepackage[pdftex, pdftitle={Article}, pdfauthor={Author}]{hyperref}
\usepackage{wasysym}
\usepackage{enumitem}
\hypersetup{colorlinks=true, citecolor=black, linkcolor=black, urlcolor=blue} 
\usepackage{ulem}

\usepackage{color}

\newcommand{\ve}[1]{\mathbf{q1}}

\newcommand{\f}{\frac}

\newcommand{\be}{\begin{equation}}      
\newcommand{\ee}{\end{equation}}      
      
\newcommand{\bef}{\begin{figure}}      
\newcommand{\eef}{\end{figure}}      
\newcommand{\bea}{\begin{eqnarray}}    
\newcommand{\eea}{\end{eqnarray}}

\newcommand{\av}[1]{\ensuremath{\left\langle q1 \right\rangle}}

\newcommand{\tve}[1]{\tilde{\boldsymbol{q1}}}

\def\bse{\begin{subequations}}
\def\ese{\end{subequations}}

\def\lsim{\raise 0.4ex\hbox{$<$}\kern -0.8em\lower 0.62ex\hbox{$\sim$}} 
\def\gsim{\raise 0.4ex\hbox{$>$}\kern -0.7em\lower 0.62ex\hbox{$\sim$}}

\def\f0N{f_0^{(N)}}
\def\bec{\begin{center}}
\def\eec{\end{center}}

\DeclarePairedDelimiter\ton{(}{)}
\DeclarePairedDelimiter\qua{[}{]}
\DeclarePairedDelimiter\gra{\{}{\}}

\DeclarePairedDelimiter\mean{\langle}{\rangle}

\begin{document}

\title{Zipf's law for cosmic structures: how large are the greatest structures in the universe?} 
  
\titlerunning{Zipf's law for cosmic structures}
  
\authorrunning{De Marzo, Sylos Labini \& Pietronero}  
  
  \author {Giordano De Marzo$^{1,2,3}$, Francesco Sylos Labini$^1$, Luciano Pietronero$^{1,2,4}$}

        \institute{
          Centro  Ricerche Enrico Fermi, Via Panisperna 89 A,
           I-00184 Rome, Italy 
          \and 
    Sapienza School for Advanced Studies, ``Sapienza'', P.le A. Moro, 2, I-00185 Rome, Italy.
 \and 
          Istituto dei Sistemi Complessi (ISC) - CNR, UoS Sapienza,P.le A. Moro, 2, I-00185 Rome, Italy.   
              \and 
    Dipartimento di Fisica Universit\`a ``Sapienza",  P.le A. Moro, 2, I-00185 Rome, Italy.
 }

\date{Received / Accepted}

\abstract{The statistical characterization of the distribution of visible matter in the universe is a central problem in modern cosmology. In this respect, a crucial question still lacking a definitive answer concerns how large are the greatest structures in the universe. This point is closely related to whether or not such a distribution can be approximated as being homogeneous on large enough scales. Here we assess this problem by considering the size distribution of superclusters of galaxies and by leveraging on the properties of Zipf-Mandelbrot law, providing a novel approach which complements standard analysis based on the correlation functions. We find that galaxy superclusters are well described by a pure Zipf's law with no deviations and this implies that all the catalogs currently available are not sufficiently large to spot a truncation in the power-law behavior. This finding provides evidence that structures larger than the greatest superclusters already observed are expected to be found when deeper redshift surveys will be completed. As a consequence the scale beyond which galaxy distribution crossovers toward homogeneity, if any, should increase accordingly.}
\maketitle


\section{Introduction}
	The distribution of matter in the universe is one of the most fascinating examples of complex structures \citep{pietronero2005statistical, labini2011inhomogeneities} and it is surely the one concerning the largest objects ever observed. Indeed, such a distribution has been proven to have power-law correlations, corresponding to a fractal structure with dimension $D \approx 2$	
	 up to several tens of Megaparsecs \citep{pietronero2005statistical}. Whether or not on larger scales it presents a crossover toward homogeneity is still matter of considerable debate \citep{whitbourn2014local,labini2014spatial, conde2015fractal, alonso2015homogeneity, pandey2015testing, shirokov2016large, heinesen2020cosmological, teles2021fractal}. 
The fractal behavior of visible matter in the universe corresponds to the fact that galaxies are distributed in a hierarchical manner: they form small groups, that, in turn, aggregate into clusters of galaxies and then, going on, clusters are grouped into larger structures, i.e., superclusters and filaments \citep{de1953evidence, abell1989catalog}, which  are the largest known structures in the universe. 
 Superclusters are linked by these filaments of galaxies and clusters forming the so called cosmic web \citep{bond1996filaments} or superclusters-void network \citep{einasto1980structure}, corresponding to a complex distribution of matter characterized by large voids and connected over-densities \citep{tully2014laniakea, pomarede2017cosmic, pomarede2020cosmicflows, colin2019evidence}.

After the surprising discovery of the cosmic web, much attention has been devoted to the quantitative statistical characterization of such large scale structures, which represent a fundamental aspect of the observable universe and a key issue for any cosmological theory. Indeed standard cosmological theories are based on the assumption of a uniform distribution of matter and so the identification of the "end of greatness" of galaxies structures represent a key issue of observational astrophysics. In the present paper we show that there is no statistical evidence that the available data contain the largest existing structures. Additionally to this, large scale structures carry important information about the primordial universe and pose intriguing theoretical problems concerning their formation \citep{einasto2007superclusters, oort1983superclusters}. For these reasons many authors have studied the features of the cosmic web, considering the distribution of voids \citep{hamaus2014universal, pan2012cosmic, einasto2011towards} and that of superclusters \citep{einasto1997supercluster} and have tried to reproduce this complex structure with N-body numerical simulations of theoretical models \citep{springel2005simulations}.

	In the present work we analyze the large scale structure of the universe considering the size distribution of superclusters of galaxies and leveraging on the properties of Zipf-Mandelbrot law \citep{Mandelbrot, zipf1949human}, which may provide interesting hints on the greatest size of these objects. Previous analyses have considered Zipf's scaling for voids \citep{gaite2005zipf, tikhonov2006properties}, which were found to be well described by this statistical law, and the probability distribution of superclusters size \citep{chow2014two}, which was 
	found to be a power-law. It is worth noting that cosmological N-body simulations do not show such a power law behavior \citep{chow2014two} and consequently they miss this feature of the large scale structure of the universe. On the other hand, to the best of our knowledge, Zipf-Mandelbrot law for galaxy structures has never been considered. As we discuss below, the first problem to be considered is that of defining a reliable list of supercluster's size: we will do it by using different definitions and trying to assess the robustness of the results on different {{observational} catalogs. {It is worth pointing out that in the present work no analysis on catalogs obtained by N-body simulation is performed.} In particular, we try to quantify the deviations from a pure Zipf's law, which represent one of the key elements in this statistical characterization of structures. Indeed,	it has been recently shown such deviations allow one to characterize the upper cutoff of the power law probability distribution of structures size \citep{de2021dynamical}. It is interesting to note that 
	such cutoff is indirectly related to the scale up to which the distribution of matter in the universe can not be approximated as being statistically uniform, providing a complementary quantitative tool beyond that based on the correlation functions analysis.

\vspace{1.5em}

The paper is organized as follows:
	 In Section II we shortly recall the definition of Zipf's law and Zipf-Mandelbrot law, pointing out how the level of deviations from Zipf's law can be used to obtain a characterization of the upper cutoff of the power law probability distribution. Then in Section III we present the superclusters catalogs used in the analysis and we compute Zipf-Mandelbrot parameters for all of them. 
Section IV is devoted to the discussion of these results, which are also compared with those obtained for clusters catalogs. A measure of statistical fluctuations is introduced ad applied to the various samples. 
	 Finally Section V is devoted to our final remarks and conclusions.
%


\section{Zipf's and Zipf-Mandelbrot's laws}
	\begin{figure*}[t!]
	    \centering
	    \includegraphics[width=0.95\linewidth]{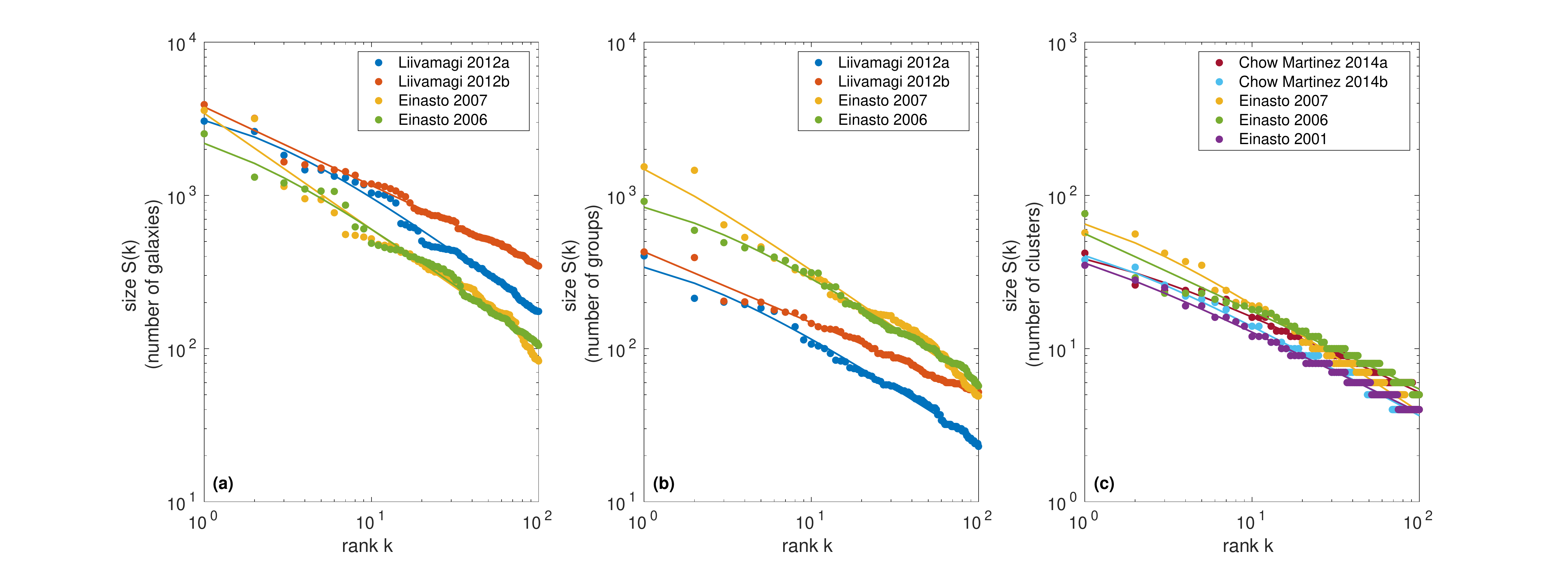}
	    \caption{\textbf{Rank-size plots of superclusters.} a) Rank-size plots of superclusters ordered by the number of galaxies they contain. b) Rank-size plots of superclusters ordered by the number of groups they contain. c) Rank-size plots of superclusters ordered by the number of clusters they contain.{Solid lines are fits to Zipf-Mandelbrot law, see the Appendix for details.} In all three cases we observe a pure Zipf's law with only minor deviations, as also discussed in the text.}     
	    \label{fig:all_rank_size}
	\end{figure*}
	Zipf's law \citep{zipf1949human} is a ubiquitous scaling law which can be considered as a sort of footprint of complexity. Indeed, this law has been observed in many complex socio-economic systems, such as cities, firms and natural language, as well as in a large number of natural systems, such as earthquakes, solar flares and lunar craters \citep{Zipf, Newman, Batty}. Many attempts have been devoted to 
	 identify a mechanism capable of explaining all these manifestations of Zipf's scaling \citep{Multiplicative, SOC, SSR, Loreto}, but a universally accepted generating process is still missing. 
	 
	 Given a system composed of $N$ objects and denoting by $S(k)$ the size of the $k$-th largest one, Zipf's law reads
	\begin{equation}	
		S(k)=\frac{S(1)}{k^{\gamma}}
	\end{equation}
	where $k$ is the rank and $\gamma$ is called the Zipf's exponent. In the case we are going to consider $S(1)$ will be the size of the largest supercluster, $S(2)$ that of the second largest one and so on. 
 Zipf's law is generally visualized by the so called rank-size plot, obtained plotting the ordered sequence of the sizes as a function of their position in the sequence: if the rank-size plot is a straight line in the log-log scale, then  the system is said to follow Zipf's law. 
	
	 Zipf's law is intimately related to power law probability distributions. Let us consider a set of objects whose sizes $S$ are power law distributed, that is 
		\begin{equation}
		\label{PS}
		P(S)=\frac{c}{S^{\alpha}},
	\end{equation}
	where $c$ is a normalization constant; the intrinsic upper and lower cutoffs of the distribution, are called respectively as $s_m$  and $s_M$ and Eq.~\eqref{PS} is satisfied for $s_m \le S \le s_M$. Note that such cutoffs are always present in real systems due to intrinsic limits on the size that an object can have: for instance, a supercluster can not contain less than a galaxy or more than all the existing galaxies. Generally a power law behavior is found only in the tail of the distribution, so in correspondence of the largest objects. It can be proven \citep{Batty, de2021dynamical} that if the underlying probability distribution is a power law, then Zipf's law is valid asymptotically in the rank, i.e. 
	\begin{equation}
	\label{eq_zipf} 
		S(k)\sim k^{-\gamma} \ \text{for}\ k\ \text{sufficiently large}
	\end{equation} 
	and moreover the Zipf's exponent $\gamma$ is related to the exponent $\alpha$ of the distribution $P(S)$ by the following expression \citep{li2002}
	\begin{equation}
		\gamma=\frac{1}{\alpha-1}.
		\label{eq:gamma_alpha}
	\end{equation}
	
	 Real systems often show deviations from Zipf's scaling law \citep{Batty, de2021dynamical}, which can be found at low ranks (for the largest objects) or at high ranks (in correspondence of the smallest elements). The latter may derive from the fact that the distribution has a  power law  behavior only in the tail or from selection effects. For instance, we could observe only a fraction of  small superclusters due to a selection effect in apparent luminosity. 
	 
	 Differently, deviations at low ranks can carry important information about the systems considered and are usually described in terms of  Zipf-Mandelbrot's law \citep{Mandelbrot} which reads 
	\begin{equation}
		S(k)=\frac{\bar{S}}{(k+Q)^{\gamma}}.
		\label{eq:zipf_mandelbrot}
	\end{equation}
	For $Q=0$ Zipf's law is recovered, while for $Q\gg 1$ strong deviations from the simple power-law behavior of Eq.~\eqref{eq_zipf} at low ranks are found. Deviations at low ranks are related to the level of sampling of the distribution and to the presence of cutoffs: indeed, as shown by \citep{de2021dynamical}, the parameter $Q$ satisfies
	\begin{equation}
	 	Q=N\ton*{\frac{s_m}{s_M}}^{1/\gamma},
	 	\label{eq:Q}
	\end{equation}
	where $N$ is the number of objects in the sample considered.
	The larger is the extent of the distribution, given by the ratio between the two cutoffs, the smaller is $Q$, while the larger is $N$  (i.e., the finer is the sampling) the larger is $Q$. Note that Zipf's law may arise in two very different ways:
	\begin{enumerate}
		\item the absence of deviations can be related to the fact that the sample available is too small to allow the detection of the intrinsic upper cutoff of the distribution. This occurs, for instance, by analyzing the size distribution of earthquakes on small time scales (tens of years) \citep{de2021dynamical}. In this case, gradually enlarging the sample leads to a growth of deviations from a pure Zipf's law;
		\item some systems spontaneously evolve out of equilibrium toward Zipf's law, as it occurs for cities \citep{de2021dynamical}. In this case, even considering the whole sample composed of all the urban settlements of a given country, deviations from Zipf's law are absent. 
	\end{enumerate}
	In both cases, whenever Zipf's law is found in a sample consisting of only a fraction of the whole system, the {\it observed} maximum cannot be used as an estimate of the {\it intrinsic} upper cutoff since it is not possible to determine if the low level of sampling is an intrinsic property of the system or an effect produced by the limited system's size. Indeed, as shown in \citep{de2021dynamical}, the {\it intrinsic} upper cutoff of the probability distribution $s_M$ can be expressed as function of the  {\it observed} maximum $S(1)$ as 
	 \begin{equation}
	 	s_M=S(1)\ton*{\frac{1+Q}{Q}}^{\gamma}.
	 	\label{eq:s_M_S(1)}
	 \end{equation}
	For $Q\to\infty$ the {\it intrinsic} upper cutoff coincides 
with the {\it observed} maximum (that is, the largest observed object), while for $Q\to0$ the upper cutoff diverges, meaning that the sample available is not sufficiently large to infer it.


\section{Zipf's law for superclusters of galaxies}
		\begin{table*}
			\centering
			\begin{tabular}{|| c || *{3}{c|}| *{3}{c|}| *{3}{c|}| }
    			\hline
				dataset    & \multicolumn{3}{c||}{N. of galaxies} & \multicolumn{3}{c||}{N. of groups} & \multicolumn{3}{c||}{N. of clusters}  \\
    			\hline
				  &  $Q$ & $Q_{min}$ & $\gamma$ & $Q$ & $Q_{min}$ & $\gamma$ & $Q$ & $Q_{min}$ & $\gamma$\\
    			\hline
				Liivamagi 2012a  &  $1.68$ & $2.21\cdot 10^{-5}$ &  $0.79$ &   $1.48$  & $2.34\cdot10^{-6}$ &   $0.70$  &  $-$ &   $-$ &  $-$ \\
				Liivamagi 2012b  &  $2.16\cdot10^{-6}$ & $8.28\cdot 10^{-8}$ & $0.51$ &  $1.54\cdot10^{-5}$  & $7.73\cdot10^{-8}$ &   $0.46$  &  $-$ &   $-$  & $-$  \\
				Einasto 2007  &  $3.12\cdot10^{-5}$ & $4.41\cdot10^{-8}$ & $0.76$ &   $0.42$  & $1.23\cdot10^{-7}$  &  $0.77$  &  $1.10$ &  $6.71\cdot 10^{-5}$  & $0.73$   \\				
				Einasto 2006 &  $1.12$ & $5.91\cdot10^{-7}$ & $0.78$ &   $1.46$ & $3.56\cdot10^{-6}$ &   $0.70$  &  $0.03$ &  $1.09\cdot10^{-7}$ & $0.51$   \\
				Einasto 2001  &  $-$ & $-$ & $-$ &   $-$  & $-$ &  $-$  &  $0.51$ &  $8.18\cdot10^{-7}$ & $0.53$   \\
				Chow Martinez 2014a   &  $-$ & $-$ & $-$ &   $-$  & $-$ & $-$  &  $0.99$ & $1.44\cdot10^{-6}$ & $0.51$   \\
				Chow Martinez 2014b   &  $-$ & $-$ & $-$ &   $-$  & $-$ &   $-$  &  $0.84$ & $1.12\cdot10^{-5}$  & $0.60$   \\
    			\hline
			\end{tabular}
		\caption{\textbf{Zipf-Mandelbrot parameters.} Deviation parameter $Q$ and Zipf's exponent $\gamma$ for all the rank-size plots shown in Fig.~\ref{fig:all_rank_size}. We also reported the minimum value of $Q$. $Q_{min}$ at $95\%$ confidence. Since all $Q_{min}$ are very close to zero, all samples are compatible with a power law with an \textit{intrinsic} upper cutoff much larger than the observed maximum.}     
        \label{tab:fits}
	\end{table*}
	Superclusters of galaxies are the largest objects in the observable universe and they can extend for hundreds of Megaparsecs containing thousands of galaxies. In order to study the rank-size relation of superclusters, first of all we have to define a suitable measure of their size. Moreover, if we want to prove that Zipf's law is a robust feature of the distribution of matter in the universe, such scaling law should be found independently of the size definition adopted. For these reasons, following the hierarchy of cosmic structures, we considered three different definitions of size: 
	\begin{itemize}
		\item number of galaxies in the supercluster;
		\item number of groups in the supercluster;
		\item number of clusters in the supercluster.
	\end{itemize}
	The main problems of these catalogs are their completeness and the possibile artifacts due to observational selection effects and clustering algorithms: both are not simple to be taken into account. In order to verify that the properties we have detected are intrinsic property of superclusters, we have  analyzed several catalogs, which have been built by using different procedures. 
		
		In the following subsections we discuss in detail the analyses we have performed.


	\subsection{Number of galaxies}
		\label{subsec:gal}
		The rank-size plots of superclusters, ordered according to the number of galaxies they contain, is reported in Fig.~\ref{fig:all_rank_size}a. Four different catalogs have been used: Einasto at al 2006 \citep{einasto2006vizier, einasto2006luminous}, Einasto et al 2007 \citep{einasto2007superclusters} and Liivamagi et al 2012 \citep{liivamagi2012sdss} (which contains two distinct catalogs corresponding to different clustering procedures). We also reported the corresponding fits to Zipf-Mandelbrot law, so to determine the parameter $Q$ and the extent of the deviations from Zipf's law. {For a detailed explanation of the fitting procedure and of the technique used for determining the uncertainty of the fit parameters see the Appendix.} As it is possible to see, superclusters follow Zipf's law with negligible deviations at low ranks.


	\subsection{Number of groups}
		Groups are the first level of aggregation of galaxies, for instance the Milky Way is part of the Local Group, which contains approximately $70$ galaxies. The catalogs we used for studying the rank-size plot of superclusters ordered by the number of groups they contain are the same we considered in the previous subsection. The rank-size plots and the corresponding fits to Zipf-Mandelbrot law are reported in Fig.~\ref{fig:all_rank_size}b: also in this case only small deviations from Zipf's law can be detected.
		

	\subsection{Number of clusters}
		Finally groups of galaxies merge into clusters, which are the level of aggregation just below superclusters. In order to study the adherence to Zipf's law when superclusters  are measured through the number of clusters they contain, we used five catalogs: Einasto et al 2001 \citep{einasto2001optical}, Einasto et al 2006 \citep{einasto2006vizier, einasto2006luminous}, Einasto et al 2007 \citep{einasto2007superclusters} and Chow-Martinez et al 2014 \citep{chow2014two} (which contains two distinct catalogs). The rank-size plots and the corresponding fits to Zipf-Mandelbrot law are reported in Fig.~\ref{fig:all_rank_size}c and again the adherence to Zipf's law is almost perfect.


	\section{Discussion}

			\begin{figure*}[t]
		    	\centering
		        \includegraphics[width=0.9\linewidth]{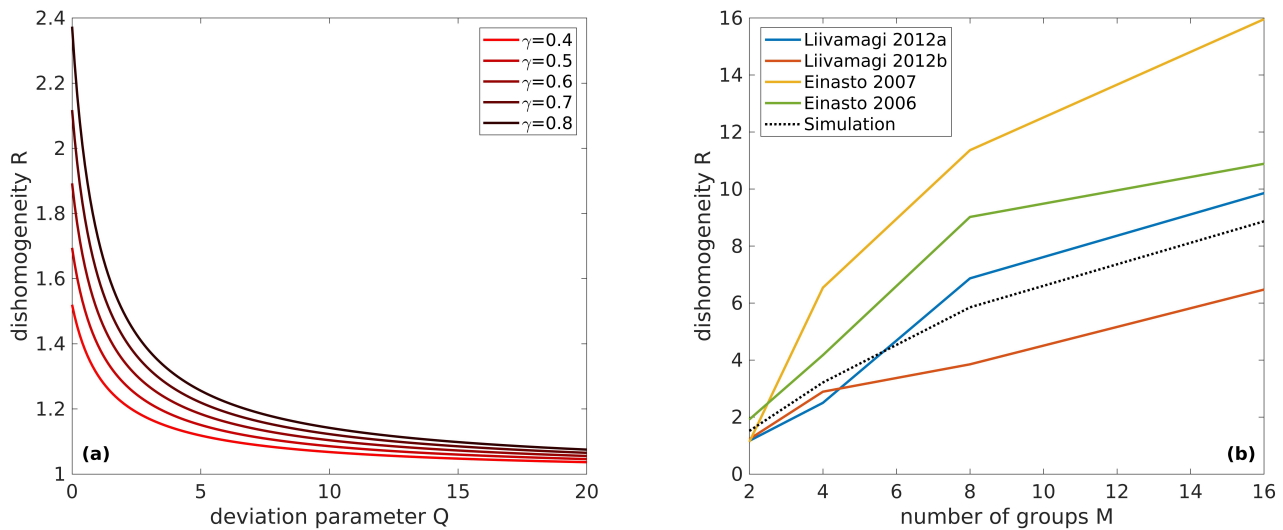}
		        \caption{\textbf{Inhomogeneity} \textbf{(a)} Average large scale fluctuations of two subsamples as function of the deviation parameter $Q$ and for various values of Zipf's exponent $\gamma$. The index $R$ is defined as the ratio between the size of largest structure in the first subsample and that of the largest structures in the second subsample. In this way $R=1$ corresponds to the absence of large scale fluctuations. \textbf{(b)} Fluctuations of the four catalogs we analyzed in Fig.~\ref{fig:all_rank_size}a when they are divided in $M$ subgroups. In this case the ratio is computed between the size of the largest structure and the size of largest structure of subsample $m$, where $m$ is the subsample with the smallest largest structure (compared to the other subsamples). {The dotted line is the fluctuation parameter of a synthetic sample of $100$ sizes which perfectly adhere to Zipf's law with exponent $\gamma=0.6$.}}    
		       	\label{fig:dishomogeneity}
			\end{figure*} 
			
			In the previous subsection we have analyzed the rank-size distribution of superclusters of galaxies considering different definitions of their size and different catalogs. In all cases negligible deviations from Zipf's law have been observed. Results are summarized in Tab.~\ref{tab:fits}, where we also report the minimal value of the deviation parameter $Q_{min}$ with $95\%$ confidence computed using a parametric bootstrap method (see Appendix for details). In all cases the minimum value of $Q$, i.e. $Q_{min}$, is almost null, meaning that the system we considered is well described by a perfect Zipf's law, regardless of the definition of size adopted or of the catalog analyzed. As discussed above this implies that the available data are not sufficient to determine the {\it intrinsic} upper cutoff of superclusters distribution, since, as shown in Eq.~\eqref{eq:s_M_S(1)}, this cutoff diverges for $Q\to0$. In other words, the conclusion is that the {\it observed} maximum is a bad estimator of the {\it intrinsic} one.	
		 This suggests that cosmic structures that are much larger than those contained in the catalogs we analyzed, are likely to be found when larger portions of the universe will be available. Moreover, the absence of a cutoff implies that the distribution of matter in the universe can not be approximated as homogeneous on the scales of the surveys considered. Indeed, even if a truncation point should exist, the distribution of superclusters is de facto scale invariant up to the largest scales currently observed. We discuss the implications of these results with respect to the standard analysis of homogeneity in the next section.


		\subsection{Inhomogeneity of the universe}
			Standard statistical methods based on the determination of the reduced
		two-point correlation function, or of its Fourier conjugate, are not suitable 
		to test whether a certain distribution is homogeneous in a given sample. The reason is that these statistical tools compare the amplitude of fluctuations to the sample density, i.e. they assume the sample density to be an unbiased estimator of the density. Such a situation occurs only if the distribution is homogeneous and here it comes the problem of testing whether or not a distribution is homogeneous without assuming a-priori that this is verified inside a given sample. This point has been discussed at length, e.g. in \citep{pietronero2005statistical, labini2011inhomogeneities} and we refer the interested reader to those works 
		for a more detailed discussion of the normalization problem.
		
		There is an additional and more subtle problem which may affect any statistical 
		determination in a finite sample, i.e. even those in which the sample density is an unbiased estimator of the {\it intrinsic}  average density. Indeed, 
		statistical analyses of finite sample distributions usually assume that fluctuations are self-averaging, meaning that they are statistically similar in different regions of the given sample volume. By determining the conditional density, i.e. the average galaxy density around a galaxy,  
		\cite{labini2009breaking, labini2009absence} have tested whether this assumption is satisfied in several sub-samples of the Sloan Digital Sky Survey. The results was that the probability density function (PDF) of conditional fluctuations, filtered on large enough spatial scales (i.e., $r>30$ Mpc/h), shows relevant systematic variations in different sub-volumes of the survey. Instead for scales $r<30$ Mpc/h the PDF is statistically stable, and its first moment presents scaling behavior with a negative exponent around one. Thus, while up to $30$ Mpc/h galaxy structures have well-defined power-law correlations, on larger scales it is not possible to consider whole sample average quantities as meaningful and useful statistical descriptors of intrinsic statistical properties. The conclusion of these studies was that this situation is due to the fact that galaxy structures correspond to density fluctuations which are too large in amplitude and too extended in space to be self-averaging on such large scales inside the sample volumes: galaxy distribution is thus inhomogeneous up to the largest scales, i.e. $r \approx 100$ Mpc/h, probed by the SDSS samples. It is worth pointing out that the self-averaging property is a necessary, but not sufficient, condition for the distribution of matter to be homogeneous. Indeed, as aforementioned, even if up to $30$ Mpc/h galaxy structures are self averaging, they still present power law correlations and so a fractal behavior, that is all but homogeneous.

		As we are going to show in what follows, the lack of the self-averaging property is related to the absence of deviations from Zipf's law for superclusters. Indeed, as we pointed out, Zipf's law corresponds to an under-sampled probability distribution and consequently, splitting the sample, worsen the level of sampling. This implies that moving from subsample to subsample, strong fluctuations are expected. In order to show this we introduce a measure of the large scale statistical fluctuations of the sample, $R$. Given a sample divided in $M$ subsamples, this parameter is defined as 
			\begin{equation}
				R(M)=\frac{\underset{m}{\max\ }\gra*{s_m(1)}}{\underset{m}{\min\ }\gra*{s_m(1)}},
				\label{eq:dishomogeneity_R}
			\end{equation}
			where $s_m(1)$ is the size of the largest element contained in the $m$th subsample. For $R=1$ all subsample present similar large scale structures, while for $R> 1$ fluctuations appear. 
			
			\begin{figure*}[t]
		    	\centering
		        \includegraphics[width=0.9\linewidth]{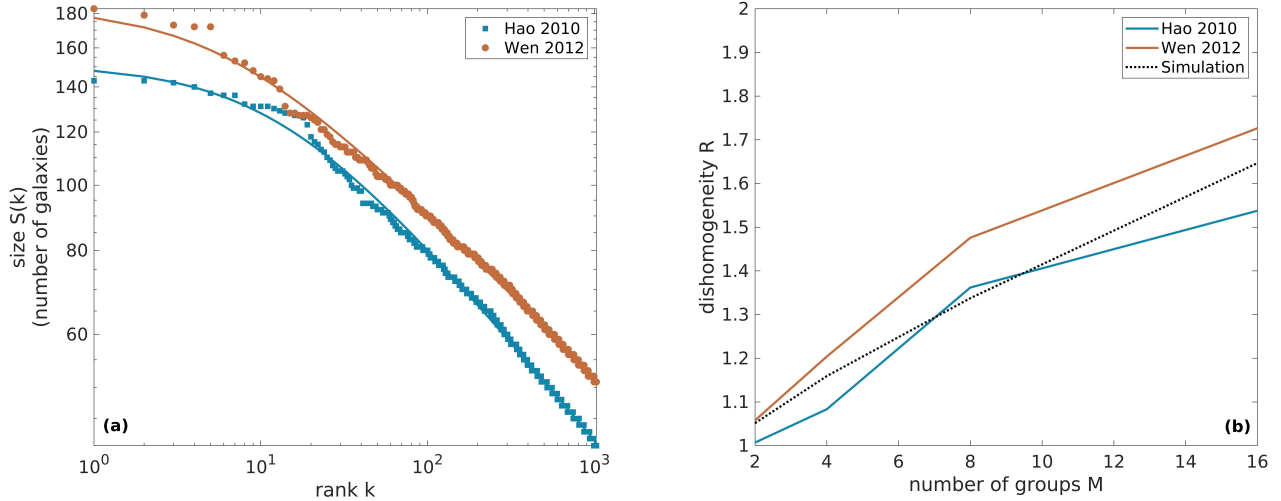}
		        \caption{\textbf{Clusters of galaxies.} a) Rank-size plots of galaxies clusters measured by the number of galaxies they contain. In this case there are strong deviations from Zipf's law for both the catalogs we considered. b) Large scale statistical fluctuations parameter $R$ of the two catalogs as function of the number of groups $M$, even for large $M$ the parameter $R$ is close to one. This implies that no strong fluctuations are observed. 
		        {The black dotted line is the fluctuation parameter of a synthetic set of $N=10^3$ elements which adhere to Zipf-Mandelbrot law with parameters $\gamma=0.27$, $Q=5$ and no statistical noise.}}    
		       	\label{fig:clusters_galaxies}
			\end{figure*} 
			
			First of all let us consider the case $M=2$, which corresponds to splitting the sample in two halves, and let us suppose that the system obeys to Zipf-Mandelbrot law Eq.~\eqref{eq:zipf_mandelbrot}, as done by the one we are studying. The size of the $k$-th largest structure satisfies 
			\[
				S(k)=\frac{\bar{S}}{(Q+k)^{\gamma}}
			\]
			and, as a consequence, it holds 
			\[
				\begin{cases}
					\text{Prob}\qua*{R=\ton*{\frac{2+Q}{1+Q}}^{\gamma}}=\frac{1}{2}\\
					\text{Prob}\qua*{R=\ton*{\frac{3+Q}{1+Q}}^{\gamma}}=\ton*{\frac{1}{2}}^2\\
					\vdots\\
					\text{Prob}\qua*{R=\ton*{\frac{n+Q}{1+Q}}^{\gamma}}=\ton*{\frac{1}{2}}^{n-1}\\
				\end{cases}
			\]
			The mean fluctuation parameter $\mean*{R}$ is then
			\[
				\mean*{R}=\sum_{n=2}^{\infty}\ton*{\frac{n+Q}{1+Q}}^{\gamma}\ton*{\frac{1}{2}}^{n-1}=2\sum_{n=2}^{\infty}\ton*{\frac{n+Q}{1+Q}}^{\gamma}\ton*{\frac{1}{2}}^n.
			\]
			This quantity is plotted as function of $Q$ and for different values of $\gamma$ in Fig.~\ref{fig:dishomogeneity}a. As it is possible to see, the larger is $Q$, the larger is the level of sampling of the inherent distribution, the lower is the level of fluctuations. The values of $\gamma$ measured in the catalogs are in the range $\gamma\in [0.4, 0.8]$; then for $Q=0$, i.e., for a perfect Zipf's law, we have $1.5<R<2.5$. This implies that one subsample will contain, on average, a structure that is between $50\%$ and $150\%$ larger than the largest structure of the other set. We then conclude that in absence of deviations from Zipf's law the two subsamples are consistently different when looked at the scale of superclusters.

			Things worsen if we divide the sample in $M>2$ groups, as shown in Fig.~\ref{fig:dishomogeneity}b. In this case we plot the parameter $R$ determined in the four catalogs we analyzed in Subsec.~\ref{subsec:gal} for increasing values of $M$, that is when the sample is divided in more and more subsets. We also report the results for a simulated sample following Zipf-Mandelbrot law with $\gamma=0.6$ and $Q=0$.  It is possible to see that the larger is $M$ and the larger is $R$; in particular, for $M=16$, the large scale statistical fluctuations parameter $R$ ranges from $6$ \citep{liivamagi2012sdss} to $16$ \citep{einasto2007superclusters}. Note that this implies that if we split the observed distribution of matter in $16$ sub-samples, one of them will contain on average a structure up to $16$ times bigger that the largest supercluster contained in the other subsamples. This indicates that at the scale of superclusters, so at scales of about one hundred Megaparsecs, the distribution of matter is not self-averaging and that consequently it can not be approximated as being statistically uniform. {In the same figure we also plotted (black dotted line) the fluctuation parameter $R$ obtained for a synthetic sample composed of $N=100$ structures which perfectly adhere to Zipf's law with exponent $\gamma=0.6$ and in absence of any statistical noise.}


		\subsection{Clusters of galaxies}

			A natural issue to investigate is the behavior of the distribution of matter in the universe when looked at smaller scales. This can be done, for instance, by analyzing the rank-size distribution of clusters instead of that of superclusters.  To this aim, we considered two large datasets: the first is described in \cite{hao2010gmbcg} and the second in \cite{wen2012catalog}. The corresponding rank-size plots and the corresponding fits to Zipf-Mandelbrot law are reported in Fig.~\ref{fig:clusters_galaxies}a: it is possible to see that there are strong deviations from Zipf's law. The  deviations parameters $Q$ with their minimum value at $95\%$ confidence bound, together with Zipf's exponents $\gamma$, are reported in Tab.~\ref{tab:clusters_param}. In both cases $Q$ is not compatible with a null value, meaning that at the level of aggregation of clusters, the cutoff of the distribution is clearly visible and the distribution itself is completely sampled. Consequently we do not expect to observe clusters much richer than those already observed when larger portions of the universe will be mapped. Moreover, as discussed above, the presence of strong deviations from Zipf's law also implies that at the scales of clusters, i.e., on scales of few Mpc, the distribution of matter is self averaging, confirming previous findings \citep{labini2009breaking, labini2009absence}. Indeed, as shown in Fig.~\ref{fig:clusters_galaxies}b, the inhomogeneity parameter $R$, Eq.~\eqref{eq:dishomogeneity_R}, is much smaller than that obtained in the case of superclusters, Fig.~\ref{fig:dishomogeneity}b, being less than $2$ when the sample is divided into $M=16$ groups.
			{We also plotted (black dotted line) the fluctuation parameter of a synthetic sample composed on $N$ elements described by a Zipf-Mandelbrot law with parameters $\gamma=0.27$, $Q=5$ and in absence of any statistical noise.} Again, note that this do not imply that on small scales the distribution of matter is homogeneous, but only that it is self-averaging. Indeed, previous studies have shown that galaxies structures are fractal on the scale of galaxy clusters.
							\begin{table}[t]
				\centering
				\begin{tabular}{|| *{4}{c|}|}
	    			\hline
					dataset &  $Q$ & $Q_{min}$ & $\gamma$ \\
	    			\hline
					Hao 2010 &  $13.28$ & $8.96$ & $0.30$\\
	    			\hline
	    			wen 2012 &  $6.13$ & $2.39$ & $0.25$\\
	    			\hline
				\end{tabular}
				\caption{\textbf{Clusters of galaxies.} Zipf-Mandelbrot parameters for the two catalogs of clusters we considered, Hao et al 2010 and Wen et al 2012.}
				\label{tab:clusters_param}
			\end{table}	
			
			
	\section{Conclusions}
		The determination of the properties of the distribution of matter in the universe is one of the most important and challenging problems in modern cosmology, since the presence of large scale structures of galaxies may be in contrast with the ``end of greatness''. This distribution shows power-law correlation, corresponding to a fractal behavior with dimension $D \approx 2$ up to tens of Mpc; it is thus natural to asses statistical properties of such a distribution using the tools of complex systems. Many works have gone in this direction after the finding that matter in the universe is characterized by a complex hierarchical structure rather than being spread in a statistically uniform way; here we present a novel approach to the problem based on Zipf's law. In particular we focus on galaxies superclusters, the largest structures in the universe, assessing their adherence to Zipf-Mandelbrot law. 
		
		Considering several catalogs and different definitions of the supercluster size, we found out that superclusters show Zipf-Mandelbrot law with  deviation parameter $Q$ close to zero, meaning that they almost perfectly adhere to Zipf's law. This finding has several implications:
		\begin{itemize}
			\item despite the catalogs we considered are the largest available, they do not show the presence of an intrinsic upper cutoff of the probability distribution of superclusters, that is consequently scale free up to the sizes of the largest structures observed;
			\item the absence of an intrinsic cutoff in the available samples implies that structures much larger than those currently observed are expected to be found as soon as deeper catalogs will be released;
			\item the probability distribution of superclusters is only poorly sampled and, as a consequence, different subsamples of the catalogs we considered show different large scale properties. This implies that the distribution of matter is not self averaging on the scales of tens of Mpc, confirming previous findings, and thus that it can not be approximated as being homogeneous on such scales.
		\end{itemize}
		
		Exploiting the same methodology we also analyzed the distribution of matter on smaller scales, considering the distribution of clusters rather that that of superclusters. In this case strong deviations from  Zipf's law are observed and therefore, on scales of few Mpc, the distribution of matter is self averaging, confirming previous findings which have also shown this distribution to be fractal on such scales. Our study shows that the self-averaging property is deeply entangled with the level of sampling of the probability distribution of cosmic structures and with the presence of deviations from  Zipf's law. This approach thus provides a novel and complementary vision with respect to those based on the study of the correlation functions or other statistical properties. Moreover it also demonstrates that Zipf's law, often considered as a mere empirical law, can be used to gather fundamental information about systems characterized by a complex structure. The application of this Zipf's law based analysis to the forthcoming galaxy surveys {and to N-body simulated catalogs} represent a natural and interesting extension of the study presented in this work. {In particular, by following the dynamical evolution of the structures during a numerical simulation, it would be possible to determine if Zipf's law is an intrinsic features of superclusters of galaxies or a spurious manifestation of this scaling law \citep{de2021dynamical}.}
	{
	\begin{acknowledgements}
		We thanks the referee Professor Bernard Jones for his helpful comments about the manuscript and for his suggestions about possible further studies.
	\end{acknowledgements}	
	}

		\bibliographystyle{aa}

\section*{Appendix}
		\subsection*{Derivation of the expressions for $Q$ and $s_M$}
			As in \cite{de2021dynamical}, we consider a truncated power law distribution of sizes, $P(S)$, that is
			\begin{equation}
			    	P(S)=
				\begin{cases}
					0 \ \text{for} \ S<s_m \\
					\frac{c}{S^{\alpha}} \ \text{for} \ s_m\leq S \leq s_M \\
					0 \ \text{for} \ S>s_M
				\end{cases}
				\label{eq:trunc}
			\end{equation}
			where $c$ is the normalization constant, and $s_{m}$ and $s_{M}$ are the lower and upper cutoff of the distribution. We recall that in the present case $S$ coincides with the size of superclusters. These cutoffs are connected to $c$ by the normalization condition
			\begin{equation}
				c\int\limits_{s_m}^{s_M}\frac{ds}{s^{\alpha}}=1 \ \rightarrow \ c=\frac{\alpha-1}{s_m^{1-\alpha}-s_M^{1-\alpha}}
			\label{eq:c}
			\end{equation}
			Now, consider the PDF $P(S)$ of a continuous variable $S$, the values of its Cumulative Distribution Function (CDF) $C(S)$ are equiprobable. This because the CDF is defined as $C(S)=\int_{s_m}^S ds'\,P(s')$ and so, performing the change of variable from $S\to C=C(S)$ and denoting $f(C)$ the PDF of $C$, we get $f(C)=\frac{dS(C)}{dC}P(S)|_{S=S(C)}=1$ for $0\le C\le 1$. As a consequence, given $N$ values of $S$ independently extracted from $P(S)$, with good approximation they can be taken as uniformly spaced in the corresponding variable $C$. Thus, the $k^{\mbox{\small{th}}}$ size ranked value $S(k)$ approximately corresponds to the CDF value  $\frac{N+1-k}{N+1}$. In formulas
	        \[
	            \int\limits_{s_m}^{S(k)}P(S)dS=c\int\limits_{s_m}^{S(k)}\frac{ds}{s^{\alpha}}\simeq\frac{N+1-k}{N+1}\,,
	        \]
	       which, together to Eq.~\eqref{eq:c}, gives
	        \[
	            \frac{S(k)^{1-\alpha}-s_m^{1-\alpha}}{s_M^{1-\alpha}-s_m^{1-\alpha}}\simeq\frac{N+1-k}{N+1} \,.
		    \]
		    Making the assumption $N+1\approx N$, $s_M\gg s_m$, and introducing $\gamma=\frac{1}{\alpha-1}$, we end up with the final rank-size formula  
	        \begin{equation*}
	    	    S(k)=\qua*{\frac{Ns_m^{\frac{1}{\gamma}}s_M^{\frac{1}{\gamma}}}{Ns_m^{\frac{1}{\gamma}}+ks_M^{\frac{1}{\gamma}}}}^{\gamma}=\frac{N^{\gamma}s_m}{\qua*{k+N\ton*{\frac{s_m}{s_M}}^{\frac{1}{\gamma}}}^{\gamma}}\,.
	        \end{equation*}
	        By comparing this expression with Zipf-Mandelbrot law Eq.~\eqref{eq:zipf_mandelbrot}
	        \begin{equation}
	         	S(k)=\frac{\bar{S}}{(k+Q)^{\gamma}},
	         	\label{eq:Zipf_Mandelbrot_met}
	        \end{equation}
	        it follows
	        \begin{equation}
	    	    \begin{cases}
	    	        \gamma=\frac{1}{\alpha-1}\\
	    		    \bar{S}=N^{\gamma}s_m\\
	    		    Q=N\ton*{\frac{s_m}{s_M}}^{\alpha-1}\,
	    	    \end{cases}\label{eq:gamma_barS_Q_met}
	        \end{equation}
			These expressions relate the number of values$/$objects and the parameters of the PDF $P(S)$ on one side, and the Zipf-Mandelbrot parameters on the other. Note that $Q$ not only quantifies deviations from Zipf's law, but also quantifies the level of sampling of the inherent distribution. Indeed $Q$ is:
			\begin{itemize}
				\item the larger the wider is the statistical sample, so the larger is the numerosity of the sample $N$;
				\item the smaller the wider is the extension of the truncated power law, given by the ratio between the upper cutoff and the lower one.
			\end{itemize}		
			Finally, combining Eqs.~\eqref{eq:Zipf_Mandelbrot_met} and \eqref{eq:gamma_barS_Q_met} we obtain an expression for the upper cutoff $s_M$
			\[
				S(1)=\frac{N^{\gamma}s_m}{(Q+1)^{\gamma}}=s_M\frac{N^{\gamma}\frac{s_m}{s_M}}{(Q+1)^{\gamma}}=s_M\ton*{\frac{Q}{Q+1}}^{\gamma},
			\]
			which yields
			\begin{equation*}
				s_M=S(1)\ton*{\frac{Q+1}{Q}}^{\gamma}.
				\label{eq:s_M_met}
			\end{equation*}
			
		\subsection*{Fitting procedure and uncertainty assessment} 
			We adopted a standard non linear least squares (NLS) fitting procedure to determine the parameters of Zipf-Mandelbrot law. This tecnique, if applied to the rank-size plot (or, equivalently, to the complementary cumulative distribution), has an accuracy which is comparable to maximum likelihood estimates \citep{white2008estimating}, while being much simpler \citep{burroughs2001upper} if the upper cutoff is unknown. In particular, we used Eq.~\eqref{eq:zipf_mandelbrot} partially linearized through logarithms
		\[
			\log S(k)=-\gamma \ln\ton*{k+Q}+c,
		\] 
		where $Q$, $\gamma$ and $c$ are free parameters. The fitting algorithm returns a $95\%$ confidence bounds on these parameters, however such uncertainty does not take into account statistical fluctuations that are encountered considering different samples generated by the same power law distribution. For this reason we exploited a parametric bootstrap so to obtain a more realistic measure of uncertainty. In particular we adopted the following procedure
	\begin{itemize}
			\item we compute the parameter $Q$ of the empirical data with the NLS technique;
			\item we use Eqs.~\eqref{eq:gamma_alpha} and \eqref{eq:Q} to determine the parameters of the underlying power law distribution;
			\item we generate $M=1000$ Monte Carlo samples with numerosity $N$ as the empirical sample using the power law distribution obtained in the previous step;
			\item each synthetic sample $m$ is fitted with the NLS technique, so to obtain the parameters $Q^m$;
			\item the empirical distribution of $Q$, $P(Q)$, is obtained performing an histogram of the $M$ values $Q^m$
			\item starting from the probability distribution $P(Q)$, the confidence bound for $Q$ is easily obtained using the cumulative distribution and determining the interval containing $95\%$ of the probability.
	\end{itemize}	
\end{document}